\documentclass[10pt]{iopart}

\usepackage{graphicx}

\begin{document}

\paper[Influences on bijel formation]{Influence of particle composition and thermal cycling on bijel formation}

\author{K~A~White\dag, A~B~Schofield\dag, B~P~Binks\ddag\ and P~S~Clegg\dag \S}

\address{\dag\ SUPA School of Physics and Astronomy, University of Edinburgh, Edinburgh, EH9 3JZ, United Kingdom}

\address{\ddag\ Surfactant \& Colloid Group, Department of Chemistry, University of Hull, Hull, HU6 7RX, United Kingdom}

\ead{\S\ pclegg@ph.ed.ac.uk}

\begin{abstract}
Colloidal particles with appropriate wetting properties can become very strongly trapped at an interface between two immiscible fluids. We have harnessed this phenomenon to create a new class of soft materials with intriguing and potentially useful characteristics. The material is known as a bijel: bicontinuous interfacially-jammed emulsion gel. It is a colloid-stabilized emulsion with fluid-bicontinuous domains. The potential to create these gels was first predicted using computer simulations. Experimentally we use mixtures of water and 2,6-lutidine at the composition for which the system undergoes a critical demixing transition on warming. Colloidal silica, with appropriate surface chemistry, is dispersed while the system is in the single-fluid phase; the composite sample is then slowly warmed well beyond the critical temperature. The liquids phase separate via spinodal decomposition and the particles become swept up on the newly created interfaces. As the domains coarsen the interfacial area decreases and the particles eventually become jammed together. The resulting structures have a significant yield stress and are stable for many months. Here we begin to explore the complex wetting properties of fluorescently-tagged silica surfaces in water-lutidine mixtures, showing how they can be tuned to allow bijel creation. Additionally we demonstrate how the particle properties change with time while they are immersed in the solvents.
\end{abstract}

\section{Introduction}
\label{intro}

Emulsions can be permanently stabilized by a layer of colloidal particles trapped at the liquid-liquid interface~\cite{Binks06}. Since both the colloids and the emulsion droplets are mesoscale objects the emulsions can have quite different properties to those stabilized by conventional surfactants. For example, the significant area that the colloid eliminates from the boundary between the two liquids leads to a deep energy minimum. Controlling colloidal emulsification is usually a question of controlling the wetting properties of the particles: partial wettability by both fluids is a pre-condition for trapping on the interface. Systematically varying the wettability can lead to an emulsion inverting from water-in-oil to oil-in-water, a process called transitional inversion~\cite{Binks00}, which demonstrates that a relationship exists between the wettability of the particle surfaces and the subsequent curvature of the droplets. Interfaces crowded with trapped colloids have solid character: in response to an elevated surface pressure they will crumple rather than eject particles~\cite{Aveyard00}~\cite{Vella04}~\cite{Clegg08}. 

An intriguing scenario arises when emulsions are stabilized using neutral wetting particles (i.e. a wetting angle close to 90$^{\circ}$); the liquid-liquid interfaces have the potential to accommodate many changes in the sign of the mean curvature. It transpires that it is possible to create emulsions where the droplets have actually become a system of tortuous channels as was first demonstrated using computer simulations~\cite{Stratford05} and subsequently shown experimentally~\cite{Clegg07}~\cite{Herzig07}. These novel emulsions have fluid-bicontinuous domains with the interface between them covered with a solid-like layer of colloids. They are known as bijels: bicontinuous interfacially-jammed emulsion gels~\cite{Stratford05}. It is not currently possible to create the domain geometry by directly mixing the constituents; instead it is created by making use of fluid-fluid demixing via spinodal decomposition. By varying the liquid composition we have demonstrated~\cite{Herzig07} how the kinetic pathway influences the final structure: for off-critical quenches a droplet emulsion is formed with the minority phase dispersed. The fluid-bicontinuous structure is tuneable: by varying the particle volume fraction we have shown how the fluid domain size can be controlled. This scaling behaviour is supported by high-resolution confocal microscopy showing approximately a monolayer of colloids on the interface~\cite{Herzig07}. This implies that the two phases are in contact at the interstices between the particles, a situation which is useful for a broad range of applications~\cite{Patent}.

In this paper we explore the relationship between the particle composition and bijel formation in some detail. Previous studies have shown that the wetting and adsorption from water-lutidine mixtures onto silica or glass surfaces is complex. This binary fluid system has a lower critical solution temperature. For samples beginning in the single-fluid phase, lutidine adsorbs on silica surfaces leading to particle aggregation~\cite{Gladden75}~\cite{Ghaicha88} (in a related system lutidine layer formation has been resolved~\cite{Privat94}). Maximum adsorption occurs from solutions containing only a small proportion of lutidine, i.e. the lutidine concentration is roughly half that of a critical composition sample; this adsorption behaviour could be related to a prewetting transition~\cite{Ghaicha88}~\cite{Beysens85}. Although not the maximum, the adsorption from critical composition samples is still significant. Surprisingly, once the phase boundary to the two-fluid phase is crossed, it is the water-rich phase that wets either silica or glass~\cite{Ghaicha88}~\cite{Pohl82}~\cite{Amara93}. The wetting transition temperature is likely to be highly sensitive to the preparation of the surfaces~\cite{Amara93}; moreover, it may be the case that not all adsorbed lutidine is displaced at the transition~\cite{Zoungrana95}. Above the wetting transition temperature the glass or silica surfaces studied exhibit partial wettability with the water-rich and lutidine-rich phases. These results indicate that tuning the wetting properties of this system is going to be challenging and that the resulting behaviour may be temperature dependent. 
\begin{figure}
\centerline{\includegraphics[scale=0.09]{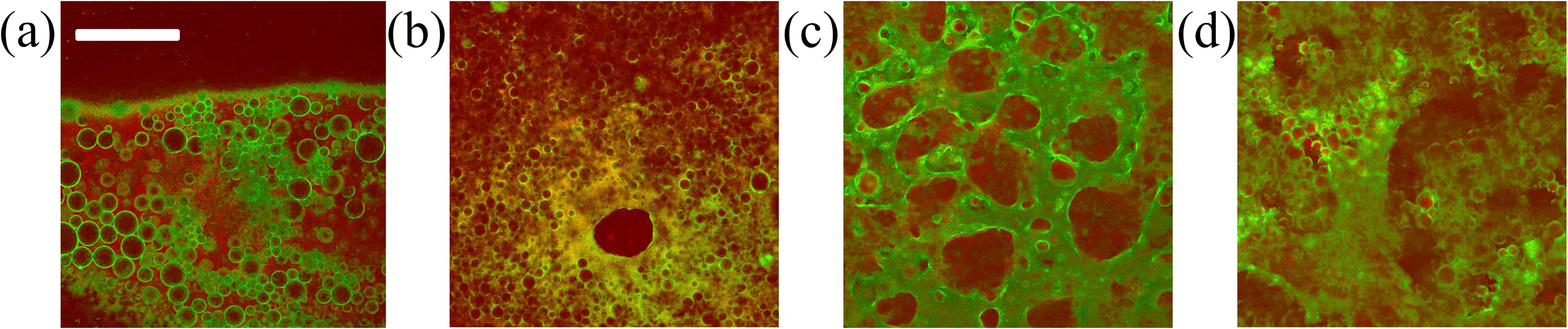}}
\caption{\label{Transitional_inversion_slices}
Confocal microscopy images of critical composition water-lutidine emulsions stabilized using each of the four types of silica particle: (a) core-shell (no surface dye) (b) 1 batch of dye (c) 1.5 batches of dye (d) 2 batches of dye. The emulsions were created in cuvettes and were prepared by warming at 17$^{\circ}$C\,min$^{-1}$ to 40$^{\circ}$C. Samples (a) and (b) are comprised of water-rich droplets in a lutidine-rich continuous phase, sample (c) is a bijel and sample (d) consists of lutidine-rich droplets in a water-rich continuous phase. Scale bar; 200\,$\mu$m.}
\vspace{1.5cm}
\centerline{\includegraphics[scale=0.15]{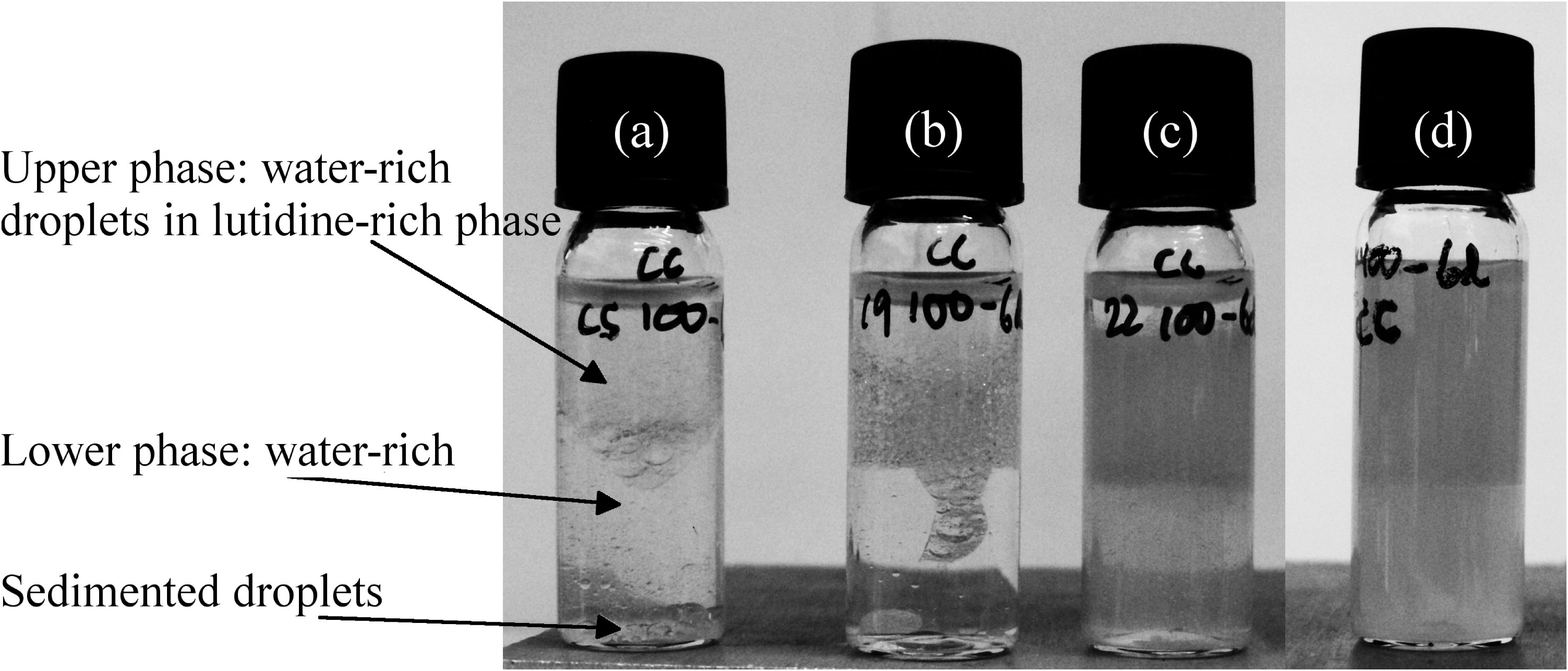}}
\caption{\label{Transitional_inversion_Late}
Images of vials (in an incubator at 40$^{\circ}$C) of critical composition water-lutidine emulsions stabilized using each of the four types of silica particle: (a) core-shell (no surface dye) (b) 1 batch of dye (c) 1.5 batches of dye (d) 2 batches of dye. The systems are shown after they have been left to stand. The upper phase of samples (a--b) are water-rich droplets in a lutidine-rich phase. The upper phase of sample (c) consists of small water-rich droplets which sediment within the lutidine-rich phase. In (a--c) the lower phase is water-rich and contains a small proportion of sedimented droplets. Sample (d) is lutidine-rich droplets in a water-rich continuous phase.}
\end{figure}

A related interplay of adsorption, wettability and emulsification has recently been investigated in the context of combinations of silica particles and cationic surfactants~\cite{Binks05}~\cite{Binks07a}~\cite{Lan07}~\cite{Binks07b}. The cases explored involved silica particles which initially have a pronounced preference for the water-phase of an oil--water system. By adding low concentrations of cationic surfactant it is possible to modify the wettability of the particles such that they form a stable emulsion. By studying particle stability, zeta potential, contact angles and emulsion lifetimes it has been shown that the cationic surfactants adsorb in a conformation where the charged head-group neutralizes a charged site on the particle surface and the hydrocarbon tail is exposed making the surface increasingly hydrophobic~\cite{Binks05}. At higher surfactant concentrations various new effects emerge including destabilization of the emulsion~\cite{Lan07} and double emulsion inversion~\cite{Binks07b}. 

Here we will reiterate our route to creating bijels. We will go on to show new results exploring the effect of modifying the surface chemistry of the colloidal particles on bijel formation using this route. We will also describe our use of mechanical agitation to create droplet emulsions using a range of particle compositions. The results deepen our understanding of bijel formation while demonstrating that, on the molecular level, there is still much to be understood.

\section{Methods and Materials}
\label{Methods}
The binary fluid pair is 2,6-lutidine (2,6-dimethylpyridine)--water~\cite{Grattoni93} which has a lower critical solution temperature (T$_c$~=~34.1$^{\circ}$C, mole fraction of lutidine x$_l$~=~0.064), where warming leads to phase separation via spinodal decomposition into domains of roughly equal volume. A critical composition mixture has pH~$\approx$~10. The particles are monodisperse St\"{o}ber silica~\cite{Stober68}; the surface chemistry is modified (and visibility enhanced) via fluorescent tagging with fluorescein isothiocyanate (FITC) using a similar procedure to ref.~\cite{Imhof99}: To make the dye, 0.0714~g of fluoroscein isothiocyanate was mixed with 0.3654~g 3-(aminopropyl)triethoxysilane for 24 hours in the dark. The resultant slurry was mixed with 2.6~ml of anhydrous ethanol to form one batch of dye. The silica particles were made by placing 1.5~l of ethanol and 186~ml of a 35\% solution of ammonium hydroxide in water into a 2.5~l flask in a fridge at 10$^{\circ}$C where the mixture was allowed to cool for 3~hours. In a separate flask 60~ml of tetraethoxysilane (TEOS) was also cooled to 10$^{\circ}$C.  After 3~hours the TEOS was added to the ethanol/ammonia mixture and immediately afterwards the required batches of dye introduced. The reaction mixture was kept at 10$^{\circ}$C and allowed to react for 24~hours. The resultant particles were cleaned by washing five times with distilled water. Here we examine particles with four different concentrations of FITC (including zero concentration): fluorescein-labelled silica core, with pure silica shell (r$\approx$400~nm); labelled with 1 batch of dye (r$\approx$230~nm) labelled with 1.5 batches of dye (r$\approx$380~nm) and labelled with 2 batches of dye (r$\approx$370~nm). A more modest increase in size with dye concentration was observed in reference~\cite{Imhof99}. The particles were carefully dried before use, an important step~\cite{Amara93}~\cite{White08}.

The bijels are prepared by initially dispersing silica particles (2\,vol\%) in water (Milli-Q, 18\,M$\Omega$cm) using an ultrasound probe, 7-8\,W for 2\,min; the 2,6-lutidine (Sigma Aldrich, redistilled, 99+\%) is only added once the particles are dispersed in water and re-cooled. This gives a suspension of colloids in the single-fluid phase. The samples need to be warmed at a controlled rate while minimizing temperature gradients. This is carried out in a temperature-stabilized aluminium block which is shaped to hold a glass cuvette (Starna Scientific) of 1\,mm pathlength. The block is pre-heated to 40$^{\circ}$C and then the suspension-filled cuvette is placed in the block; this gives a warming rate of 17$^{\circ}$C\,min$^{-1}$. Mixtures for the bulk emulsion studies were made in glass sample vials in a similar manner to that described above, except that 0.3-0.5 vol\% of particles was used. Given that the particle batches have different radii, the proportion added was adjusted so that each sample contained particles to cover roughly the same interfacial area. Vials were placed in the incubator at 40$^{\circ}$C and separation occurred.  Repeated shaking of the warmed vials gave reproducible behaviour.

In recent confocal microscopy studies rhodamine B dye, (Fluka, standard) was added to a concentration of $7.5 \times 10^{-6}$~M, to create a distinction between the two fluid domains. The dye preferentially partitions into the lutidine-rich domain and can be excited at 543~nm. The particles (excited at 488~nm) emit at 525~nm while the rhodamine emits at 572~nm. The confocal microscope (Nikon TE300) is used in conjunction with a Biorad Radiance 2100 scanner with an Argon-ion laser and a green helium-neon laser. To image the sample within the aluminium block we use an extra-long working distance $\times$20 PlanFluor Nikon objective with an adjustable correction collar.
\begin{figure}
\centerline{\includegraphics[scale=0.6]{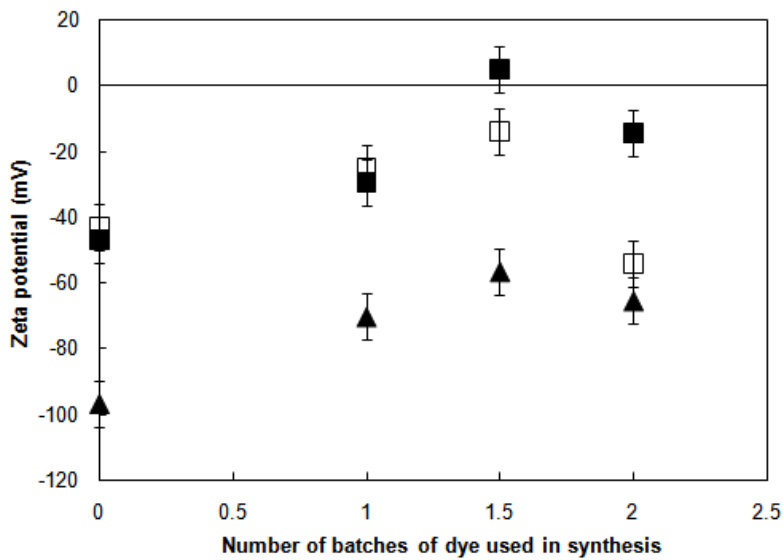}}
\caption{\label{Zeta_Potential} Zeta potential values for the four batches of particles used in this study under various conditions. The full squares are the values for particles dispersed in distilled water taken after one hour, (the open squares are the values for the particles when they have been left in distilled water overnight) and the full triangles are the values when the pH is raised to 10.}
\end{figure}

Zeta potential measurements were performed using a Malvern Zetasizer 3000~HS. Samples were prepared by dispersing 0.01\,wt\% particles in one of the following solvents, again using a sonicating probe at 7-8\,W for 1-2\,min.: (a) distilled water (pH $\approx$ 6), with mobility measurements collected after an hour; (b) distilled water, with measurements collected after waiting overnight and (c) distilled water adjusted to pH~=~10, with measurements collected after an hour.

\section{Results}

\subsection{Particle composition and bijel formation}

Figure~\ref{Transitional_inversion_slices} shows images of a series of emulsions prepared with the four batches of particles described in section~\ref{Methods} using the standard route to bijel creation described therein.  The lutidine-rich phase appears red while the particles appear green. A systematic change in the emulsions formed is observed for the different batches of colloids. For samples (a) and (b) a colloid-stabilized emulsion has formed where the droplets are dark, i.e. water-rich. This behaviour is typical if the particles have a slight preference for the lutidine-rich phase. For sample (d) the emulsion droplets are red inside. This indicates that they are lutidine-rich droplets, suggesting that these particles have a preference for the water-rich phase. Between these extremes is sample (c) where convoluted red and dark domains are entangled with each other separated by a percolating interface; this arrangement is characteristic of the bijel structure. Across the sequence of samples the concentration of dye used during synthesis increases and this appears to be associated with an increase in the particles' preference for the water-rich phase.

Figure~\ref{Transitional_inversion_Late} shows samples of similar composition to Figure~\ref{Transitional_inversion_slices} but created in vials (see section~\ref{Methods}), placed in an incubator at 40$^{\circ}$C and shaken in a manner to promote droplet formation. For all samples the binary fluid pair water--lutidine has close to critical composition (mole fraction of lutidine x$_l$~=~0.0644~$\pm$~0.0002) and the images were taken in the incubator; the densities of the two phases are quite similar which slows down separation. The location of the particles has been identified from the yellow colour of the FITC (there is no rhodamine B in these samples); the water-rich phase is the more dense and where possible it will sink. We observe that the particles are towards the top of all of the vials although this belies a more complex spectrum of behaviour. This is most obviously shown by the \textit{dripping droplets} of vial (b). The upper phases of samples (a) and (b) are comprised of large water-rich droplets stabilized by colloidal particles. These droplets are slowly sagging down under the weight of the water (see also Figure~\ref{Transitional_inversion_slices}(a) where there is a larger, black, outer region of the water-rich phase). Water-rich droplets would be expected for colloidal particles that are preferentially wetted by the lutidine-rich phase which is consistent with Figure~\ref{Transitional_inversion_slices}(a) and (b). The flat interface between the water-rich (lower) and lutidine-rich (upper) phases is presumably coated with a layer of particles. Sample (d) in Figure~\ref{Transitional_inversion_Late} is comprised of lutidine-rich droplets creaming within a water-rich continuous phase. This is what is expected for interfacial colloids that are preferentially wetted by the water-rich phase (see also Figure~\ref{Transitional_inversion_slices}). Sample (c) has the composition of fluids and the treatment of particles for bijel formation. For this mixing route the resulting emulsion consists of small water-rich droplets which sediment within the lutidine-rich phase. Small droplets are anticipated close to the inversion point. The particle composition required for bijel formation is very close to that which induces transitional inversion to take place when emulsification is via shaking. These observations, for a completely different emulsion preparation route, further support the link between increasing dye concentration and particles with surfaces possessing an increasing preference for the water-rich phase.
\begin{figure}
\centerline{\includegraphics[scale=0.3]{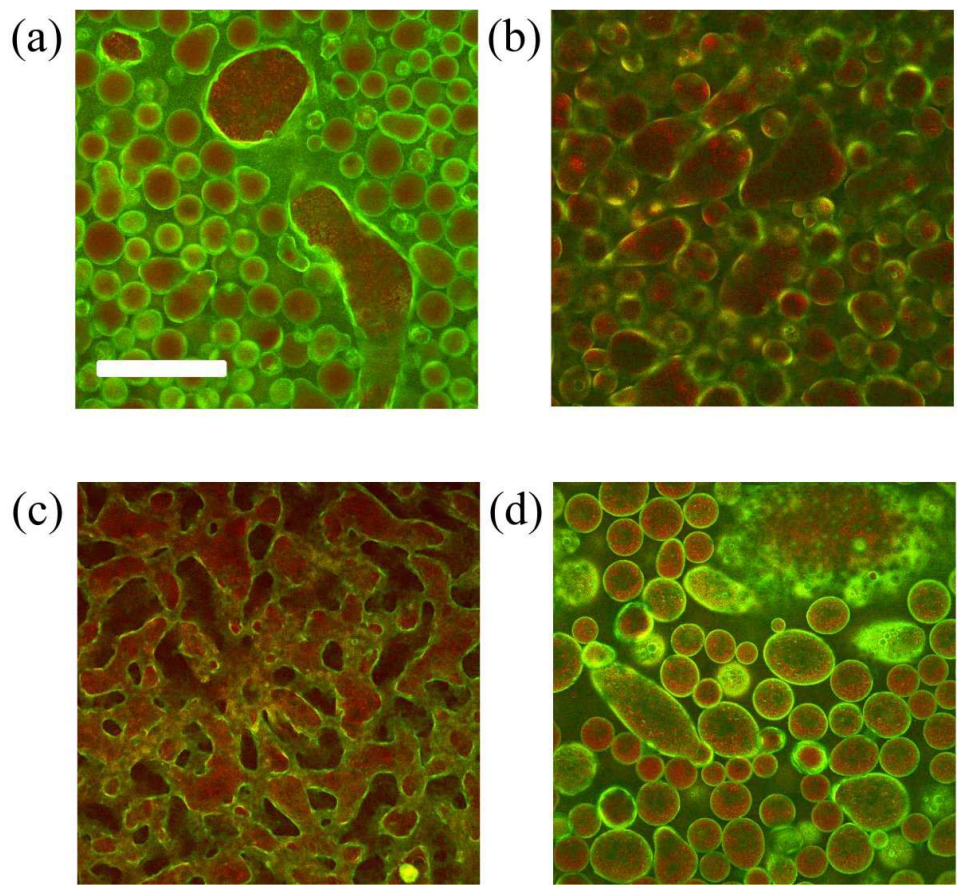}}
\caption{\label{Shaking_bijels}
Confocal microscopy images of colloid-stabilized emulsions at 40$^{\circ}$C prepared by various routes (full details are given in the text). The lutidine-rich phase appears red while the particles appear green. (a) A droplet emulsion prepared by warming from room temperature to 40$^{\circ}$C; the particles have a slight preference for the water-rich phase; (b) a droplet emulsion prepared by shaking the sample shown in (a); (c) a bijel made by cooling the sample shown in (b) re-dispersing the particles at room temperature and warming from room temperature to 40$^{\circ}$C; (d) a droplet emulsion prepared by shaking the sample shown in (c). The two warming cycles were carried out at a rate of 17$^{\circ}$C min$^{-1}$. Scale bar; 200\,$\mu$m.}
\end{figure}

To show some of the changes on the particle surfaces zeta potential measurements were made, see Figure~\ref{Zeta_Potential}. Initially we will consider just the measurements made after one hour in distilled water and those made at pH 10 (chosen since this is the pH of critical composition water-lutidine mixtures). The zeta potential of the core-shell silica (no surface dye) in distilled water is of the same order as that seen for similar particles~\cite{Wu06}. With increasing dye concentration the surface potential tends to become less negative in water; although there is some deviation from this trend. By contrast, raising the pH makes the zeta potentials substantially more negative --- independent of the dye concentration used in synthesis. Considered in concert with Figure~\ref{Transitional_inversion_slices} and Figure~\ref{Transitional_inversion_Late} these results suggests that the particles' acquiring a preference for the water-rich phase is associated with a decrease in zeta potential of their surfaces: a trend more normally associated with an increased preference for a low dielectric constant solvent.

\subsection{Thermal cycling and bijel formation}

Figure~\ref{Shaking_bijels} shows a sequence of fluorescence confocal microscopy images for a single sample as a sequence of manipulations are carried out. The process illuminates some of the changes in behaviour of particles as a function of time. The first frame, Figure~\ref{Shaking_bijels}(a), shows an emulsion prepared using the standard bijel creation route from the 1.5 batch of dye particles dried~\cite{White08} so that they have a marked preference for the water-rich phase: the dispersed phase is red (i.e. lutidine-rich) and the continuous phase has a green appearance due to the population of surplus particles that prefer this phase to the interface. The next frame, Figure~\ref{Shaking_bijels}(b), shows the same sample after it has been withdrawn from the aluminium block, manually shaken, and then replaced in the 40$^{\circ}$C block. Droplets rich in lutidine have still formed; however, the substantial surplus of particles in the continuous phase has now gone. This could be due to the change in emulsification procedure or due to a change in the surface properties of the particles. After taking this image the sample was then re-cooled into the single-fluid phase and the particles re-dispersed by shaking at room temperature. From here the warming route for bijel preparation was again followed: the third frame, Figure~\ref{Shaking_bijels}(c), shows the resulting bijel. The structure has tortuous domains and interfaces showing variations in mean curvature characteristic of this structure. Note that the temperature changes leading up to frames (a) and (c) were identical --- only shaking and one thermal cycle make the results different. The change appears to be due to the exposure of the particles to the liquids in the intervening period. The final frame, Figure~\ref{Shaking_bijels}(d), shows the result of withdrawing the bijel and shaking it and replacing the cuvette in the 40$^{\circ}$C block. Lutidine-rich droplets again result, with a slight surplus of particles within the droplets. Evidently the particles now have begun to have a preference for the lutidine-rich phase although not enough to yield a complete inversion. It appears clear that with exposure to the solvents the particle surfaces begin to lose their initial preference for the water-rich phase. Some aspects of this behaviour could be associated with the warming and cooling rates that we employ: the adsorption effects described in section~\ref{intro} were studied using much slower changes in temperature and there may be a time lag before the particle interfaces reach equilibrium.

Further zeta potential studies were carried out on the particles as a function of time, as shown in Figure~\ref{Zeta_Potential}. The particles were kept in distilled water overnight after which further measurements were then made. Prolonged exposure to distilled water had very little effect on the zeta potential of particles with little or no dye on the surface. The two batches which were synthesised with higher concentrations of dye show significant time dependence. The 1.5 batch of dye sample shows a zeta potential which becomes negative again while the 2 batch of dye sample time almost has a similar effect to a substantially raised pH. In previous studies~\cite{Herzig07} we kept a bijel sample for seven months to observe its stability. The particle fluorescence was not lost in this time which may suggest that the zeta potential effect is not the result of a chemical breakdown of the dye in water. Taken together, these studies may indicate that long exposure to the solvents (especially water) acts to increase the preference of the particles for the lutidine-rich phase via the enhancement of the surface charge; this is the opposite trend to that induced by increased dye concentration.

\section{Discussion}

Our results show that bijel formation occurs for samples with close to critical composition where the colloidal particles have surface chemistry such that the system is close to the point at which transitional inversion would occur. This is not enough on its own: what is also crucial to bijel formation is the kinetic pathway~\cite{Herzig07}. Phase separation via spinodal decomposition will lead to bijel formation while simple mechanical agitation will not for this combination of fluids and particles. 

The batches of particles differ due to the addition of dye (FITC) and our observations suggest that this holds the key to the surface chemistry of the silica particles. The dye is incorporated in combination with 3-(aminopropyl)triethoxysilane (APS) which is employed in excess. The APS gives amino groups to the particle surface and the proportion of these relative to silanol groups increases as the dye concentration increases~\cite{Imhof99}. To understand the wetting properties we need to consider how the surfaces will respond to different solvents. Above pH $\approx$ 2 the silanol groups, native to St\"{o}ber silica, tend to be negatively charged (Si--O$^-$). Below pH $\approx$ 9 the basic amino groups tend to add positive charges to the surfaces (R--NH$_3^+$). When the number of amino groups matches the number of silanol groups zero surface charge can be achieved: locally the surface is charged but on average the charges cancel. The FITC itself includes a carboxyl group and a phenolic hydroxy group which will tend to increase the negative charge on the surface at any pH~\cite{vanBlaaderen92}; however, the proportion of dye is significantly less than the proportion of APS. The zeta potential results show these effects, see Figure~\ref{Zeta_Potential}. With increasing dye concentration the surface potential tends to become less negative in water (similar results have been seen for particles with only APS and no dye~\cite{Wu06}). It is surprising that the highest dye concentration and hence the higher concentration of amino groups does not lead to the most positive zeta potential in distilled water. It is possible (although unlikely) that the dye carboxyl and phenolic hydroxy groups are beginning to swing the balance in favour of a negative zeta potential. When the system is modified such that pH = 10 both silanol and charged amino groups will (further) deprotonate and the surfaces will become more negatively charged; a trend which is followed by all batches of particles independent of dye concentration. 

Our observations show that changing the dye concentration modifies the wettability of the silica surfaces (at least in the presence of lutidine). If we are to understand this we need to know how the lutidine interacts with the surfaces. St\"{o}ber silica is thought to acquire a lutidine wetting layer approaching the binodal line~\cite{Gladden75}~\cite{Ghaicha88}. Additionally the parent compound, pyridine, has been used as a test for the existence of acid groups, such as silanol, at silica and other surfaces~\cite{Zoungrana95}~\cite{Basila64}. It is reasonable to assume that adsorption via electrostatic interactions, hydrogen bonding and other mechanisms will also occur for lutidine. So how will this behaviour be modified by the incorporation of dye? The dominant effect of the dye addition procedure is to replace silanol sites with amino groups and systematically increasing the dye concentration appears to increase the preference for the water-rich phase. One possible explanation is that the lutidine plays a role on the silica surfaces similar to a cationic surfactant~\cite{Binks05}. For the core-shell particles (no surface dye) the lutidine adsorbs via electrostatic interactions or hydrogen bonding to the surface silanol groups. The number of exposed charged groups decreases and the resulting particles prefer the lutidine-rich phase. For the other batches of particles (with surface dye) the amino surface groups reduce the adsorption of lutidine leading to increasingly hydrophilic particle surfaces. This picture ignores much of the complexity of this system including the strong temperature dependence before and after phase separation.

Time-dependent behaviour is clear in the zeta potential results (Figure~\ref{Zeta_Potential}) with the magnitude of the \textit{overnight} potential growing much faster than linearly with the concentration of dye used in synthesis. The behaviour is different to the direct charging effect of raising the pH, which has an effect that is essentially independent of the dye concentration. Time-dependent behaviour is also evident in the emulsion results in Figure~\ref{Shaking_bijels}. This too may be due to exposure to water leading to the conclusion that exposure to water tends to make the surfaces increasingly wettable by the lutidine-rich phase. The chemistry underlying this behaviour is not currently understood. That this phenomena exists does, however, facilitate fine tuning of the particle wetting properties and thus is a useful handle for bijel creation.

\section{Acknowledgments}
Thanks to M.~Cates and J.~Thijssen for helpful comments. The microscopy facilities were provided by the Collaborative Optical Spectroscopy and Micromanipulation Centre (COSMIC). Funding in Edinburgh was provided by the EPSRC grant EP/D076986/1 with additional support from Scottish Enterprise POC/8-CHM-002 and the EU NoE SoftComp.

\end{document}